\documentclass[10pt]{article}
\usepackage[utf8]{inputenc}
\usepackage{graphicx}
\usepackage{amsmath,amssymb}
\usepackage[small,bf]{caption}
\usepackage[numbers,sort&compress]{natbib}

\usepackage{graphicx} 

\usepackage{geometry}
\title{Metabolic energy expenditure for time-varying isometric forces}
\author{Sriram Sekaripuram Muralidhar,$^{1}$ Kristen Heitman,$^{2}$ \\ 
Samuel C. Walcott,$^{3}$  Manoj Srinivasan$^{1,4}$ \\
$^{1}$Mechanical and Aerospace Engineering, The Ohio State University, Columbus, OH 43201 \\
$^{2}$School of Health and Rehabilitation Sciences, The Ohio State University, Columbus, OH 43201 \\
$^{3}$Mathematical Sciences, and Bioinformatics and Computational Biology,\\ Worcester Polytechnic Institute, Worcester, MA01609 \\
$^{4}$Program in Biophysics, The Ohio State University, Columbus, OH 43201}

\date{}

\begin{document}
\maketitle

\begin{abstract}
Muscles consume metabolic energy (ATP) to produce force. A mathematical model for energy expenditure can be useful in estimating real-time costs of movements or to predict energy optimal movements. Metabolic cost models developed so far have predominantly aimed at dynamic movement tasks, where mechanical work dominates. Further, while it is known that both force magnitude and rate of change of force (force rate) affect metabolic cost, it is not known how these terms interact, or if the force rate dependence can be a consequence of the force dependence. Here, we performed extensive human subject experiments, involving each subject over 5 hours of metabolic trials, which systematically changed the mean forces and forces rates  so as to characterize a holistic relation for metabolic cost based on both force and force rate --- or analogously, torque and torque rate. Our experiments involved humans producing symmetric or asymmetric sinusoidal forces  with different means, amplitudes, frequencies, and rise and fall periods. We showed that the metabolic cost can be well-approximated by a sum of power law functions of torque and torque rate. We found that the metabolic cost scales  non-linearly with joint torque  (with exponent $\gamma_1 = 1.36$) and non-linearly with torque rate (with exponent $\gamma_2 = 2.5$). Surprisingly, the data suggested that the cost  was roughly four times higher for decreasing the torque than increasing, mirroring the analogous ratio between the cost of positive and negative work. Using these metabolic cost relations, we show that if the metabolic cost scales with particular exponents with muscle force and force rates, the same exponents will be observed in multi-joint tasks with multiple muscles. Our new metabolic cost model involving both force and force rate will potentially allow better predictions of energy optimal movements and thus inform wearable robot design and analysis.
\end{abstract} 

\section{Introduction}
Muscles consume metabolic energy to produce force in isometric tasks (constant muscle length, without movement) and non-isometric  tasks (changing muscle length, with movement). Minimising metabolic energy expenditure, at least in part, predicts healthy human behaviour in tasks such as locomotion \cite{alexander1989optimization,selinger2015humans,seethapathi_metabolic_2015,brown_unified_2021,long_walking_2013}, reaching \cite{huang2012reduction,wong2021energetic}, and even in isometric force production \cite{muralidhar_metabolic_2023}. Metabolic cost is used as a metric to assess the efficacy of assistive devices such as exoskeletons and prosthesis designed for locomotion \cite{au_powered_2009,grabowski_leg_2009,collins_reducing_2015}. So, human behavioural and assistive device researchers are particularly interested in measuring the metabolic cost of various physical tasks. Metabolic energy expenditure in vivo  measured via indirect calorimetry \cite{ferrannini_theoretical_1988,brockway_derivation_1987} is time-consuming as it either requires the task to be repeated for around 5 minutes or  extrapolation of  few mins of non-steady state data  \cite{selinger_estimating_2014}, thereby limiting the number of trials performed  and  subjects studied. A mathematical model for metabolic cost can potentially be used to speed up metabolic estimation from kinematics and kinetics \cite{slade2021sensing}. Such mathematical models can also be broadly useful in  facilitating  studies of dynamic, non-repetitive or transient tasks  \cite{seethapathi_metabolic_2015}, 
 model-based design or human in-the-loop optimization of exoskeletons and prosthesis \cite{handford2018energy,collins_reducing_2015}, and simulation studies to predict energy-optimal movement behaviour \cite{kuo_simple_2001,srinivasan_fifteen_2011,ackermann_optimality_2010,miller_comparison_2014,falisse2019rapid,brown_unified_2021}.

Previous metabolic cost models have certain limitations in the prediction of metabolic cost for isometric tasks. Metabolic cost models developed from isolated muscles experiments in-vitro have generally been compared with whole body movement tasks such as walking and have not been compared in detail with in-vivo human isometric experiments designed explicitly to test those models \cite{bhargava_phenomenological_2004,houdijk_evaluation_2006,lichtwark_modified_2005,minetti_theory_1997,zahalak_modeling_1990,umberger_model_2003}. Most models are particularly deficient in predicting the cost of isometric force. For instance, the  Umberger et al model \cite{umberger_model_2003} predicted roughly 7\% of experimentally measured cost for an isomeric task involving tracking sinusoidal forces \cite{van_der_zee_high_2021}. The models developed from in vivo human experiments have either no prediction for isometric tasks as they were derived with terms specific to non-isometric tasks \cite{doke_mechanics_2005,doke_energetic_2007,dean_energetic_2011}, considered only force level changes but not time series changes  \cite{hawkins_modeling_1997, muralidhar_metabolic_2023} or only time series changes but not force level changes \cite{kuroda_electrical_1970,van_der_zee_high_2021}.

Here, we focus on developing a metabolic cost model applicable to  isometric tasks involving arbitrary time-varying force production based on joint torque and torque rate, which includes constant force as a special case. In previous work, we showed that the metabolic cost of near-constant isometric force scales non-linearly with force \cite{muralidhar_metabolic_2023}. Van der Zee and Kuo \cite{van_der_zee_high_2021} showed that force-rates have a substantial energy cost by having subjects produce forces with different frequencies. But these two studies \cite{muralidhar_metabolic_2023,van_der_zee_high_2021}  did not independently change force and force rates, so either do not have information on the cost of force rates or cannot distinguish the effect of a nonlinear metabolic cost dependence on force versus force rate.  More generally, previous in-vivo experiments usually involved univariate sweeps along some exertion parameters \cite{muralidhar_metabolic_2023,doke_mechanics_2005,doke_energetic_2007,van_der_zee_high_2021}. Here, we performed extensive human subject experiments with diverse force levels and force changes in a manner that allows us to characterize the independent contributions of force and force rate on the metabolic cost of time-varying forces. We show that a simple additive model with a nonlinear power-law cost for force and force-rates is sufficient to explain the metabolic cost of force production. Further, we examine forces with different increasing and decreasing force rates, allowing us to show that the cost of decreasing forces is higher than the cost of increasing forces. Finally, while our metabolic cost model is at the level of human joints, we provide mathematical arguments for how this joint-level model may extend to the individual muscle-level.

\section{Methods}
\begin{figure}[h!]
    \centering
    \includegraphics{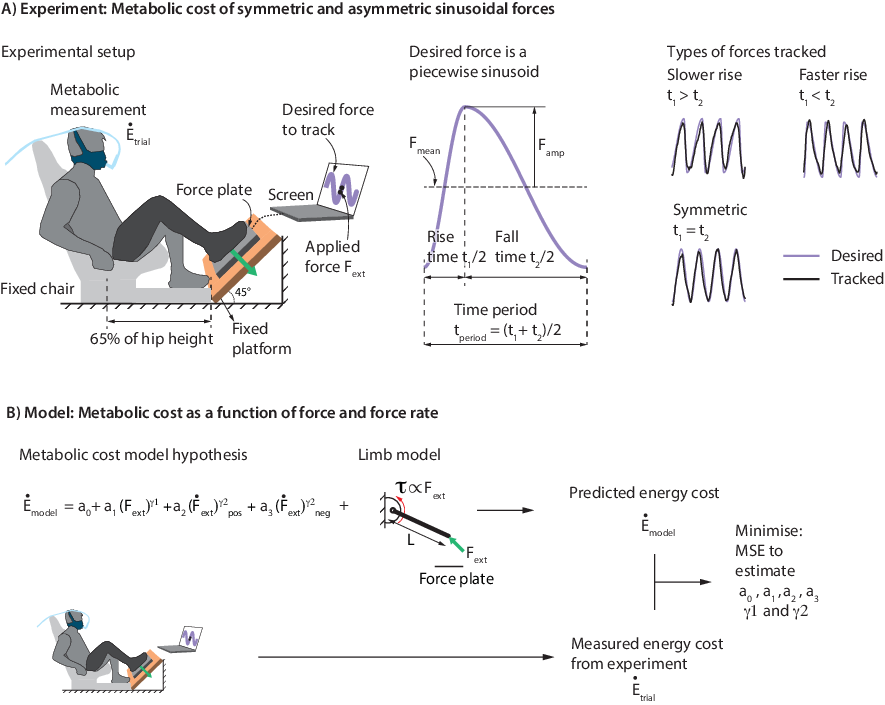}
    \caption{ \textbf{Metabolic cost experiment and modelling framework.} \textbf{A) Experiment.} Humans apply force perpendicular to the force platform while sitting still using the right leg and track a desired force displayed on the screen for 5 minutes.  The desired force is a piece-wise sinusoid formed by the combination of two sinusoids having same means (${F_\mathrm{mean}}$), amplitude (${F_\mathrm{amp}}$) but same or different rise time period ($t_1/2$) and fall time period ($t_2/2$). We then varied the means, amplitude and time periods across trials, which resulted in three broad categories of piecewise sinusoids where rise period is smaller than the fall period (${t_1 < t_2}$), rise period equals fall period (${t_1 = t_2}$) and rise period is greater than the fall period (${t_1 > t_2}$). \textbf{B) Model.} We fit a power law function of joint torque and torque rate represented in terms of externally measured force, force rate (${\dot E_\mathrm{model}}$) to the experimental metabolic cost (${\dot E_\mathrm{trial}}$). We represented the limb using a single rigid massless link with a joint and  expressed the joint torque in terms of external force measured by the force plate.  We then fit the metabolic cost model based on external force (equation \ref{eq:metabolicModel}) to the experiment data and  estimated the model parameters. } 
    \label{fig:ExperimentSchematic}
\end{figure}

\subsection{Human metabolic experiments}
We estimated the metabolic cost of producing time-varying forces from human experiments where the subjects tracked piecewise sinusoidal forces  using their right leg while sitting in a comfortable posture (Figure \ref{fig:ExperimentSchematic}A).  We measured volumetric rates of oxygen and carbon dioxide in the respiratory gases (MGC Diagnostics Corporation Ultima CardiO2 metabolic cart) and estimated the metabolic cost ${\dot{E}}_\mathrm{trial}$ using the following empirical equation \cite{brockway_derivation_1987} relating the measured quantities: ${\dot E = 16.58 \dot{V}_{CO_2}+4.51 \dot{V}_{O_2} \textrm{W/kg}}$. Each trial lasted 5 minutes, allowing us to estimate the steady metabolic cost more reliably. We measure the forces exerted by the participants perpendicular to the platform (vernier force platforms), which were streamed real-time to a computer screen (lab view DAQ NI 6008) to provide a visual feedback to the subject of the applied force as well as the desired force. The fundamental period $F(t)$ of the piecewise sinusoidal desired force was given by the following equation:
\begin{eqnarray}
    F(t) &=& F_\mathrm{mean} + F_\mathrm{amp} \cdot \sin{\left(\frac{2 \pi t}{t_1} \right)} \textrm{ for }  0 \le t < \frac{t_1}{2} \textrm{ and } \\
    F(t) &=& F_\mathrm{mean} + F_\mathrm{amp} \cdot  \sin{\left(\frac{2 \pi (t-t_1/2)}{t_2} \right)}  \textrm{ for }  \frac{t_1}{2} \le t <  \frac{t_1+t_2}{2}
\end{eqnarray}
and subsequent periods were produced by repeating this fundamental period: $F\left(t+ \, {(t_1+t_2)}/{2}\right) = F(t)$. The two sinusoids composing the piecewise sinusoid combination have the same mean and  amplitude, with potentially equal or different rise and fall time  ($t_1/2$ and $t_2/2$, Figure \ref{fig:ExperimentSchematic}A), where $t_1$ and $t_2$ are the time periods of individual sinusoids.

All experiments were approved by the Ohio State University IRB. A total of 11 subjects (7M, 4F; height = 1.67 $\pm$ 0.18 m; mass 74.45 $\pm$ 27.54 kg; age = 27 $\pm$ 8 years, mean $\pm$ 2 s.d.) participated with informed consent. Each subject performed about 30 trials split equally across two separate day sessions, with two of them completing one session each and one of them performing a different force levels for the first session. No trial was repeated across sessions or subjects, resulting in a large diverse dataset consisting of  297 unique data points across all subjects. Having different rise and fall time periods set different increasing and decreasing slopes in the sinusoid, thereby helping us to study the differences in costs of having different increasing and decreasing force rates. Piecewise sinusoidal function parameters for each trial were chosen randomly from a list containing 16 different combinations of the two sinusoid time periods such that the total time period ${t_\mathrm{period} = (t_1+t_2)/2}$ is either  1\,s, 1.25\,s or 1.5\,s. For these three time periods, the rise time parameter $t_1/2$ was respectively selected from the following three sets: \{${0.125, 0.25, 0.375, 0.5, 0.625, 0.75, 0.875\,\mathrm{s}}$\},  \{${0.25, 0.5, 0.75, 1\,\mathrm{s}}$\} and \{${0.25, 0.5, 0.75, 1.0, 1.25\,\mathrm{s}}$\}. There were three different mean force levels ($F_\mathrm{mean}$): 20\%, 35\%, 50\% of the subjects' voluntary maximum force ${F_\mathrm{max}}$, comfortable to sustain over 30\,s. The two different amplitude levels ($F_\mathrm{amp}$), 50\% and 100\% of the mean force.  Trials were drawn from this set in a random sequence and performed in the random sequence. Subjects performed 15 trials in a session with breaks in between. Before each trial, we estimated the force exerted by the subjects' passive and relaxed leg on the platform. All desired and applied forces have this passive leg force subtracted, so that the subjects had to only control the active force they exert on the platform.

\subsection{Metabolic model for time-varying forces}
 We fit a metabolic cost model based on joint torque and rate of change of torque (torque rate) with a single joint sagittal plane limb model and estimate the optimum model unknowns which minimises the mean squared error with the experiment (Figure \ref{fig:ExperimentSchematic}B). We hypothesise that the metabolic cost  is a power law function of joint torque (${\tau^{\gamma_1}}$) and torque rate (${ \dot{\tau}^{\gamma_2}}$). We further hypothesise that the metabolic cost of increasing the  torque (torque rate is positive, i.e., $\dot{\tau} >0$) is different from decreasing the torque (torque rate is negative, i.e., $\dot \tau < 0$). We used a sagittal plane limb model with a single joint and a massless rigid link to relate the joint torque with externally measured force using statics: ${\tau = \mathrm{constant}  \cdot F_\mathrm{ext}}$.  We consider a single joint limb model even though the experiment involved multi-joints because  joint torques are correlated and proportional in isometric tasks due to mechanical constraints \cite{muralidhar_metabolic_2023} and hence, for the purposes of inferring the exponents the number of joints does not matter. Without loss of generality, we used massless links to account for joint torque due to active muscle force production only, because we subtracted the passive force due to gravity from the force tasks in our experiments. We fit the following piecewise  joint torque based metabolic cost model with joint torque represented in terms of external measured force:
\begin{equation}  {\dot{E}_\mathrm{model} = a_0+a_1 F_\mathrm{ext}^{\gamma_1}+a_2 (\dot{F}_\mathrm{ext})_\mathrm{pos}^{\gamma_2} + a_3 (\dot F_\mathrm{ext})_\mathrm{neg}^{\gamma_2}}.  \label{eq:metabolicModel} \end{equation}

For each trial, we estimated the external force ${F_\mathrm{ext}}$ by averaging the time series data of the force tracked from 50 to 300\,s. We estimated force rate  ${\dot F_\mathrm{ext}}$ by using a forward difference approximation, and determined the average of the positive and negative parts of the force rates $\mathrm{(\dot F_{ext})_{pos}}$ and $\mathrm{(\dot F_{ext})_{neg}}$ over 50 to 300\,s. We averaged measured metabolic cost time series data from 180 to 300\,s  and normalised it by subject weight to estimate metabolic cost for each trial ($\mathrm{\dot E_{trial}}$). Using all 297 subject trials, we fit the metabolic cost model (equation \ref{eq:metabolicModel}) to the measured metabolic cost (${\dot{E}_\mathrm{trial}}$) to estimate  subject-specific offsets ${a_0}$ and  coefficients and exponents  (${a_1, a_2, a_3, \gamma_1, \textrm{ and } \gamma_2}$) common across all subjects by minimising the mean squared residual across all trials and subjects. We assume subject specific offsets ${a_0}$ to allow for different resting metabolic rates across subjects. In this  optimisation, we normalize the forces by 100\,N, force-rates by 100\,Ns$^{-1}$, and second derivatives of force by 100\,Ns$^{-2}$, and all expressed in these normalized units.

\section{Results}

\begin{figure}[h!]
    \centering
    \includegraphics{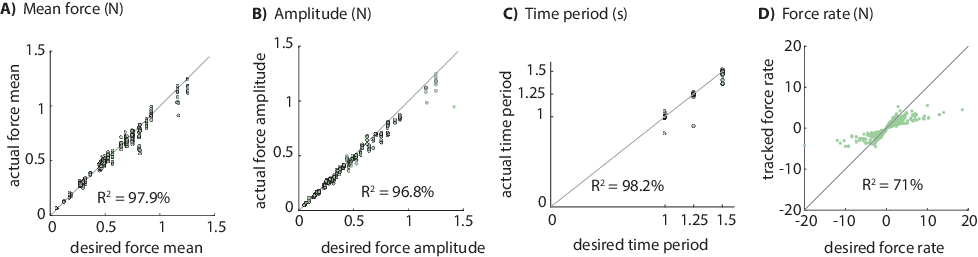}
    \caption{\textbf{Subjects' tracking of  target forces.} Tracked quantities obtained by averaging the entire 5 mins of time-series data. \textbf{A)} Piecewise sinusoid means ($\mathrm{F_{mean}}$) are tracked well by subjects. \textbf{B)} Piecewise sinusoid amplitudes ($\mathrm{F_{amp}}$) are tracked well by subjects. \textbf{C)} Piecewise sinusoid time-periods ($\mathrm{t_{period} = (t_{1} + t_{2})/2}$) are tracked well by subjects  \textbf{D)} The effective mean positive and negative force-rates across trials are imperfectly tracked by the subjects, but well piecewise sinusoid's time period ($\mathrm{t_{p}}$).} 
    \label{fig:ResultsActualVsDesired}
\end{figure}
\paragraph{Subjects tracked desired forces well-enough.} The subjects were reasonable in tracking the desired force means $F_\mathrm{mean}$, amplitudes $F_\mathrm{amp}$, and the total time period ${t_p}$ of the piecewise sinusoid ($R^2>97\%$ for all of them, Figure \ref{fig:ResultsActualVsDesired}A, B, C). They were poorer in tracking individual sinusoid rise and fall times $  {t_1/2}$ and  ${t_2/2}$ (see Appendix figure 2), subjects preferred individual time periods (${t_1}$ and  ${t_2}$) closer to symmetry than prescribed. Nevertheless, the actual rise and fall times were correlated enough with the desired times so that we obtained a big diversity of positive and negative force rates highly correlated with the desired force rates (Figure \ref{fig:ResultsActualVsDesired}D, $R^2 = $ 71\%). Because the subjects did not track the desired forces with negligible error, all calculations and model predictions use actual forces produced.

\begin{figure}[h!]
    \centering\includegraphics{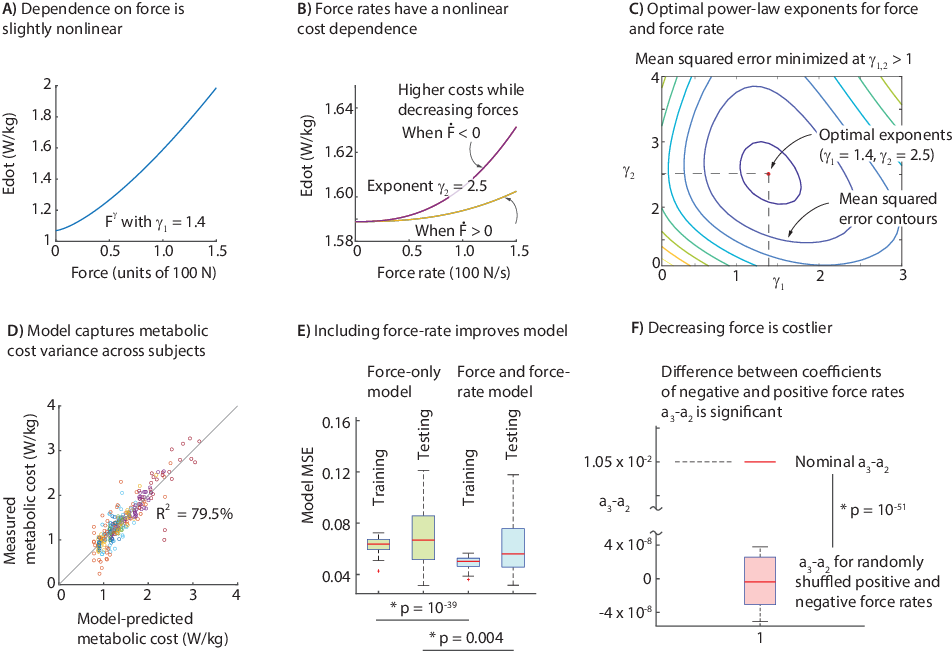}
    \caption{\textbf{Metabolic model.} \textbf{A)} Best-fit force-dependent of metabolic cost is slightly nonlinear (exponent $\gamma_1 = 1.4$). The curve shows model metabolic cost when the force is constant and force rate is zero. \textbf{B)} Best-fit force-rate-dependent of metabolic cost has a strong nonlinearity (exponent $\gamma_2 = 2.5$). Curves show model metabolic cost terms for positive and negative force-rate when the nominal force equals 100 N in our experiment. \textbf{C)} Mean squared error (MSE) of model versus experimental data is minimized at $\gamma_1 = 1.4$ and $\gamma_2 = 2.5$. \textbf{D)} The best-fit model predicts the data well, with a high $R^2$ value of 79.5\%. \textbf{E)} MSE is lower for the model having both force and force rate terms, as opposed to having only force term across training and test dataset.    \textbf{F)} The coefficients for positive and negative force rates not obtained by random chance. Repeated regression of positive and negative force rate values randomly shuffled or not with equal probability gives $a_3-a_2$ in the 10$^{-8}$ which is quite different from the nominal value.  } 
    \label{fig: ResultsOptimumModel}
\end{figure}

\paragraph{Metabolic model for time-varying forces is well-predicted by nonlinear force and force-rate terms.} We fit the metabolic cost model (equation \ref{eq:metabolicModel}) to the data. We found that the metabolic cost scales non-linearly with both force and force-rate (Figure \ref{fig: ResultsOptimumModel}A,B). The best-fit force exponent was ${\gamma_1}  = 1.36$ and the best best-fit force-rate exponent was ${\gamma_2} = 2.55$ (Figure \ref{fig: ResultsOptimumModel}A-C), achieving a R-squared value of 79.4\% (Figure \ref{fig: ResultsOptimumModel}D). We found that dropping the force-rate term and having a pure force power-law produced a strictly worse model. Using a randomly selected 80\% of the data to fit the model results in 5\% lower training error for the model including the force-rate cost ($p = 10^{-39}$); the model with the force-rate also generalizes better to the unseen 20\% of the data, again giving 5\% lower mean squared error ($p = 0.004$, Figure \ref{fig: ResultsOptimumModel}E). 

\paragraph{Decreasing force costs more than increasing force.} We found that the coefficient for negative force rate is about three times the coefficient of the positive force rate by approximately three-fold ($a_3 = 0.0155$, $p$-value = 10$^{-6}$; $a_2 = 0.0049$, $p$-value = 0.045, Figure \ref{fig: ResultsOptimumModel}B). To establish that this difference between the coefficient was statistically significant and could not be produced by random data variability, we repeated the regression with positive and negative force rate values randomly shuffled or not with equal probability. This regression with shuffled data gives $a_3-a_2$ values in the 10$^{-8}$ range, resulting in $p = 10^{-51}$ for $a_3-a_2 = 0.0105$ to be obtained by chance (Figure \ref{fig: ResultsOptimumModel}F).

\paragraph{Optimality of linear muscle force and torque scaling strategy.}
Consider a limb with multiple muscles and joints at rest, and the task was to produce an external force of different magnitudes and rates along a fixed direction while being at rest. There are usually infinitely many choices of muscle forces to produce a given external force -- due to the fact that 
the human body has more muscles crossing a joint than is minimally necessary to produce a particular torque. We find that if every muscle had a metabolic cost that had a power law form (for instance,  $F^{\gamma_1} + \dot{F}^{\gamma_2}$), a `linear scaling strategy' minimizes this power law cost summed over all muscles. That is, given the optimal muscle forces or joint torques for a given external force, any other external force in the same direction can be obtained by linearly scaling all the muscles forces or joint torques. This provides a simple solution to the 'muscle force indeterminacy problem' and simplifies. The proof of this linear scaling strategy and the linear system that provides the analytical solution for muscle forces or joint torques any external force is provided in the appendix. 

\paragraph{Whole limb cost scales the same way as muscle-level costs for static tasks.} An implication of the linear scaling strategy is that if the muscle level cost for each muscle scales like $F^{\gamma_1} + \dot{F}^{\gamma_2}$, then the whole limb cost scales in the same manner with respect to the external force ${F^{\gamma_{1}}_\mathrm{external} + \dot F^{\gamma_{2}}_\mathrm{external}}$, as shown in the mathematical appendix. This provides self-consistency of our approach and ensures that the single-segment model (figure \ref{fig:ExperimentSchematic}) initially used to obtain the metabolic cost model in terms of external force has no loss of generality.
 
\section{Discussion}
We have developed a metabolic cost model for time-varying isometric muscle contraction showing that the cost scales like a power law function  with joint torque, external force with  $\gamma_1 = 1.4$ and torque rate, external force rate with  $\gamma_2 = 2.5$. We also showed that the cost of  decreasing the torque is more than increasing.  

Our exponent for force ($\gamma_1 = 1.4$) approximately agrees with our previous result \cite{muralidhar_metabolic_2023}. More qualitatively, the nonlinear relation on force agrees with some previous in-vivo studies, which showed that oxygen consumption and the force are non-linearly related \cite{kuroda_electrical_1970-1,josenhans_muscular_1967,cerretelli_energetics_1976}. In our previous work, we estimated the force exponent from metabolic measurements of standing with knees bent and showed that this exponent predicted  both upper and lower limb force sharing better than linear or quadratic exponent \cite{muralidhar_metabolic_2023}. Our nonlinear dependence of metabolic cost on force in contrast with some previous experimental studies which suggested linear dependence of muscle metabolic cost with muscle force: but these earlier studies were either for externally activated isolated human heart or rat skeletal muscles in-vitro  \cite{paul_relation_1975,hood_oxygen_1986,gluck_aerobic_1977} or in-vivo  measurements which were performed for lower force range \cite{clarke_energy_1960,royce_oxygen_1962}. 

 Van der Zee and Kuo \cite{van_der_zee_high_2021} proposed a model of metabolic rate proportional to the second derivative of force, which is equivalent to the metabolic cost per movement being proportional to the first derivative of force. This is a different cost from our model, which they supported by showing an approximate quadratic scaling of metabolic cost with force frequency. Our model is roughly consistent with their data, also indicating a roughly quadratic with oscillation frequency, though more specifically our model predicts a faster than quadratic scaling of metabolic cost with frequency when the force mean and amplitude are fixed (${\gamma_2 > 2}$). Reviewing  Van der Zee and Kuo's data (figure \cite{van_der_zee_high_2021}) suggests their data may also be consistent with a slightly faster-than-quadratic scaling with oscillation frequency. In future work, we will consider how well alternative models with higher derivatives fit our or even more diverse data. 
   
We found that  decreasing the force is more costly than increasing the force by having different coefficients in the model for positive and negative force rate (\ref{eq:metabolicModel}). One reason positive and negative force rate may have different costs may be due to decrease force, the calcium needs to be pumped back to the sarcoplasmic reticulum which incurs a metabolic cost \cite{inoue_structural_2019,barclay_energy_2007}. This calcium pumping cost is in addition to the ATP activity that sustains repetitive actomyosin activity required for force maintenance.  At the individual muscle level,  metabolic measurements have been  performed for continuous or intermittent electrical stimulation in-vivo or in-vitro.  These studies suggest that the cost for intermittent activation is more than continuous \cite{chasiotis_atp_1987,spriet_energy_1988,hogan_contraction_1998,bergstrom_energy_1988} which is an analogous to say that the cost of producing sinusoidal force  is more than constant force.  But these studies did not perform experiments comprising different activation and relaxation times, which is analogous to having different upward and downward sinusoid slopes in our experiments.

An alternative explanation for different costs for increasing and decreasing is the use of co-contraction to reduce the output force quickly by activating the antagonist muscles to achieve the required negative force rates. Co-contraction or pre-activation of muscles is seen in a variety of ecological tasks, so increased cost of negative force rates may be behaviourally relevant.

We have shown analytically that if muscles have power law metabolic costs on force and its derivatives, such power law behaviour is inherited by whole body or whole limb tasks as seen in our experiments. Under some conditions, we showed that such whole limb force production tasks have simple optimal strategies such as linear force scaling production, which may facilitate simple yet near-optimal neural control. 

Future work could characterize dependence on joint angle or muscle length, characterize muscle activity and elastic tendon length changes during these experiments, repeat such extensive experiments for non-isometric tasks, delineate cost differences between muscles, and pursue  explaining these phenomena via multiscale models that go from molecules, through sarcomeres, muscle fibres, recruitment, connective tissue, and whole body mechanics. In conclusion, we have shown that metabolic cost scales non-linearly with joint torque and torque rate and the same exponent is applicable to muscle force sharing. Such metabolic cost models, when generalized to each joint or muscle, may be used in whole body biomechanical simulations for predictions of movement behaviour, how force is shared between muscles, and potentially for real-time metabolic monitoring \cite{annerino_towards_2022, khusainov_real-time_2013}.

\bibliographystyle{unsrt}
\bibliography{MyLibrary_april_16}

\section*{Appendix}

\subsection*{Theorem: Linear scaling strategy for muscle forces and joint torques.}
In the main manuscript, we expressed the metabolic cost as a function of  external force and force rate, using a single-link model. Now, we consider a limb with multiple joints and multiple muscles. As in our experiment, this limb at rest needs to produce a one-parameter family of external forces and force rates, all along the same direction but of different force and force-rate magnitudes. We now show that if all the muscles power-law metabolic cost, all with the same force exponent and same force rate exponent, the energy optimal muscles forces for the task follow a  `linear scaling strategy.' That is, if the energy optimal solution is known for one external force and force rate magnitude, the optimal solution for any other external force and force rate magnitude is obtained by linearly scaling all the muscle forces by one scalar factor and the force rate magnitudes by a different scalar factor. 

\begin{figure}[h!]
\centering\includegraphics{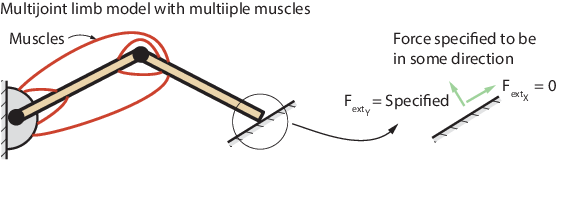}
    \caption{A multi-segment multi-joint limb with multiple muscles, applying an external force along some fixed direction. }
    \label{fig:multimuscle}
\end{figure}

The multi-joint limb has $m$ muscles (figure \ref{fig:multimuscle}A) and $p$ degrees of freedom. The limb is at rest and the task is to apply a 3D external force $\mathbf{F}_\mathrm{ext}$ of different magnitudes along a fixed direction: the external force $\mathbf{F}_\mathrm{ext}$ can be considered a scaled version $\mu \mathbf{F}_0$ of some nominal 3D external force $\mathbf{F}_\mathrm{0}  \in \mathbb{R}^3$, which decides the desired direction.  Then, force rate is given by $\mathbf{\dot F}_\mathrm{ext} = \dot \mu \mathbf{F}_0$ where $\mathrm{\dot \mu}$ is changes the force rate along the fixed direction. Because the force rate is always in the direction of the force, the force remains along a fixed direction for all time. The muscle force magnitudes are $\mathbf{F}_\mathrm{mus} = [F_1; \  F_2; \  \ldots ] \in \mathbb{R}^m$ and we minimize the metabolic cost with a power law relation on muscle force and force rate magnitudes: 
\begin{eqnarray}
     \dot{E}_\mathrm{task} &=& c_1 F_\mathrm{mus,1}^{\gamma_1} +c_2 F_\mathrm{mus,2}^{\gamma_1} + \ldots + c_m F_\mathrm{mus,m}^{\gamma_1} +d_1 \dot F_\mathrm{mus,1}^{\gamma_2} +d_2 \dot F_\mathrm{mus,2}^{\gamma_2} + \ldots + d_m \dot F_\mathrm{mus,m}^{\gamma_2} \nonumber \\ 
     &=& C^T \mathbf{F}_\mathrm{mus}^{\gamma_1} + D^T \mathbf{\dot F}_\mathrm{mus}^{\gamma_2},  \label{eq:metcostMultiJoint}  
\end{eqnarray}
where the matrix $C = [c_1;\ c_2; \ \ldots \ c_m]$, $D = [d_1;\ d_2; \ \ldots \ d_m]$,  $\mathbf{F}_\mathrm{mus}^{\gamma_1} = [F_1^{\gamma_1}; \  F_2^{\gamma_1}; \  \ldots ]$ and $\mathbf{\dot F}_\mathrm{mus}^{\gamma_2} = [\dot F_1^{\gamma_2}; \  \dot F_2^{\gamma_2}; \  \ldots ]$.

\paragraph{What is proved.} We show that if the optimal muscle force for the nominal external force and force rate ($\mu = 1$ and $\dot{\mu} = 1$) is $\mathbf{F}_\mathrm{mus0}$, the optimal muscle force for any other external force is given by $\mathbf{F}_\mathrm{mus} = \mu \mathbf{F}_\mathrm{mus0}$ and the optimal force rate is $\dot{\mathbf{F}}_\mathrm{mus} = \dot \mu \mathbf{F}_\mathrm{mus0}$.

\paragraph{Proof.} In terms of notation and proof technique, we closely follow our earlier work  \cite{muralidhar_metabolic_2023}, which proves a special case of zero force rates. The condition that the whole limb is in static equilibrium can be written as a linear equation in the list of $m$ muscle force magnitudes $\mathbf{F}_\mathrm{mus} \in \mathbb{R}^m$ as follows:
\begin{equation}  A\, \mathbf{F}_\mathrm{mus} = B\, \mathbf{F}_\mathrm{ext} =  \mu B \mathbf{F}_\mathrm{0},  \label{eq:staticsForce} 
\end{equation}
where $A$ is a $p \times m$ matrix and $B$ is a $p \times 3$ matrix containing geometric parameters, where the number of rows $p$ generally equals the number of degrees of freedom. Equations \ref{eq:staticsForce}-\ref{eq:staticsForcerate} ignores gravity, as noted earlier.
Taking the derivative of equation \ref{eq:staticsForce} gives the condition for force rates:
\begin{equation}  A\, \mathbf{\dot F}_\mathrm{mus} = B\, \mathbf{\dot F}_\mathrm{ext} = \dot \mu B \mathbf{ F}_\mathrm{0}\label{eq:staticsForcerate} \end{equation}
These equations serve as equality constraints for the optimization.
 
The Lagrangian $L$ for the constrained optimization problem is:
\[ L = C^T \mathbf{F}_\mathrm{mus}^{\gamma_1} + D^T \mathbf{\dot F}_\mathrm{mus}^{\gamma_2}+ \xi^T (A\, \mathbf{F}_\mathrm{mus} - B\, \mathbf{F}_\mathrm{ext}) + \lambda^T (A\, \mathbf{\dot F}_\mathrm{mus} - B\, \mathbf{\dot F}_\mathrm{ext}),  \]
where $\xi$ contains the Lagrange multipliers $[\xi_1;\ \xi_2; \ \ldots \ \xi_p]$ and $\lambda$ contains the Lagrange multipliers $[\lambda_1;\ \lambda_2; \ \ldots \ \lambda_p]$. Computing the gradients $\nabla_F\,L$ and $\nabla_{\dot{F}}\,L$ with respect to the unknown muscle force  $\mathbf{F}_\mathrm{mus}$ and force rate magnitudes  $\mathbf{\dot F}_\mathrm{mus}$, and setting these gradients equal to zero gives:
\begin{eqnarray}  
\nabla L &=& \gamma_1 C^T \mathbf{F}_\mathrm{mus}^{\gamma_1-1} + A^T \xi = 0 \textrm{ and } \label{eq:dLdF} \\
\nabla L &=& \gamma_2 D^T \mathbf{\dot F}_\mathrm{mus}^{\gamma_2-1} + A^T \lambda = 0. \label{eq:dLdF_dot} 
\end{eqnarray} 
Equations \ref{eq:staticsForce} and \ref{eq:staticsForcerate} together provide $2p+2m$ linear equations in the $2p+2m$ unknowns in the $\xi$ , $\mathbf{F}_\mathrm{mus}$, $\lambda$ , $\mathbf{\dot F}_\mathrm{mus}$.

We now show that these equations \ref{eq:staticsForce}, \ref{eq:staticsForcerate}, \ref{eq:dLdF}, and \ref{eq:dLdF_dot} imply the linear scaling strategy. That is, we show that the solution is of the form:
\begin{eqnarray}
    \mathbf{F}_\mathrm{mus} &=& \mu \mathbf{F}_\mathrm{mus0} \  \textrm{ and } \  \xi = \mu^{\gamma_1-1} \xi_0, \ \textrm{ and }  \label{eq:solutionformforce} \\
    \mathbf{\dot F}_\mathrm{mus} &=& \dot \mu \mathbf{F}_\mathrm{mus0} \  \textrm{ and } \  \lambda = \dot \mu^{\gamma_2-1} \lambda_0, \label{eq:solutionformforcerate} 
\end{eqnarray}  
where $\mathbf{F}_\mathrm{mus0}$ and $\mathbf{\dot F}_\mathrm{mus0}$ are a fixed set of muscle force and force rate  magnitudes respectively. To show this, we substitute this proposed solution form into equations \ref{eq:staticsForce}, \ref{eq:staticsForcerate}, \ref{eq:dLdF}, and \ref{eq:dLdF_dot}, which gives the following  three equations:
\begin{equation}
\mu A \mathbf{F}_\mathrm{mus0} = \mu B \mathbf{F}_\mathrm{0}, \  \textrm{ and } \gamma C^T \mathbf{F}_\mathrm{mus0}^{\gamma_1-1} + A^T \xi_0 = 0, \ \textrm{ and } \gamma D^T \mathbf{F}_\mathrm{mus0}^{\gamma_2-1} + A^T \lambda_0 = 0 \label{eq:withnomudependence}
\end{equation}
    
These equations (equation \ref{eq:withnomudependence}) have no dependence on $\mu$ and $\dot \mu$, and thus can be solved for $\mathbf{F}_\mathrm{mus0}$, $\xi_0$ and $\lambda_0$ without any dependence on $\mu$ and $\dot \mu$. This non-dependency of $\mathbf{F}_\mathrm{mus0}$ on $\mu$ and $\dot \mu$ is consistent with the assumed solution form (equations \ref{eq:solutionformforce} and \ref{eq:solutionformforcerate}). Thus, the proposed solution  does correctly specify how  the muscle force magnitudes and the force rate magnitudes depend on $\mu$ and $\dot \mu$ respectively, thereby establishing the linear scaling strategy.

\paragraph{Corollary 1: Muscle-level power law cost implies whole limb-level power law costs.}
Substituting the solution for muscle force and force rates (equation \ref{eq:solutionformforce}-\ref{eq:solutionformforcerate}) into the whole limb or `whole body' metabolic cost expression (equation \ref{eq:metcostMultiJoint}) gives:
\begin{align*}
\textrm{Whole limb metabolic cost} &= C^T \mathbf{F}_\mathrm{mus}^{\gamma_1} + D^T \mathbf{\dot F}_\mathrm{mus}^{\gamma_2} \\ &= C^T (\mu \mathbf{F}_\mathrm{mus0})^{\gamma_1} +D^T (\dot \mu \mathbf{F}_\mathrm{mus0})^{\gamma_2}
\\ &= \mu^{\gamma_1} \cdot C^T \mathbf{F}_\mathrm{mus0}^{\gamma_1} + \dot \mu^{\gamma_2} \cdot D^T \mathbf{F}_\mathrm{mus0}^{\gamma_2}. \label{eq:metcostscaling}
\end{align*}
Given that $\mu$ and $\dot{\mu}$ are respectively proportional to the external force and force rates, we have shown that the whole limb metabolic cost also scales in the same power law manner (same exponents) with respect to the external force and force rate as do the muscles with respect to the muscle forces and force rates.

\paragraph{Corollary 2: Linear scaling strategy is optimal for joint torques.}
At a given configuration, the joint torques are linear functions of muscle forces, given by:
\begin{equation}
\tau = D\, \mathbf{F}_\mathrm{mus} = \mu\, D\, \mathbf{F}_\mathrm{mus0} = \mu\, \tau_0, \label{eq:jointTorques}
\end{equation}
where $D$ is a matrix of muscle moment arms and $\tau_0$ is the nominal joint torques when $\mu = 1$. Equation \ref{eq:jointTorques} shows that the linear scaling strategy is also true for joint torques: joint torques for some external force are scaled versions of the joint torques for a particular external force in the same direction. Differentiating equation \ref{eq:jointTorques}, we obtain the analogous linear scaling strategy for the joint torque rates:
\begin{equation}
\dot \tau = D\, \mathbf{\dot F}_\mathrm{mus} = \dot \mu\, D\, \mathbf{F}_\mathrm{mus0} = \dot \mu\, \tau_0. 
\end{equation}

 \paragraph{Other remarks.} While we have shown the three results for metabolic costs that are functions of force and force rate, they are true for when the metabolic cost depends in similar power law fashion on higher derivatives of force as well. 

 The theorem and corollaries, as proved, rely on ignoring gravity. However, it can be shown that the theorem and corollaries are also true in the presence of gravity in the following special case: the multi-segment system is such that it can be at rest with muscles turned off while in contact with the external surface. The 2D two segment system shown in figure \ref{fig:multimuscle} has this property. While such a system is in rest with such passive turned-off muscles, there is no metabolic cost, but there is an external force applied on the surface, $\mathbf{F}_\mathrm{passive}$. In this situation, all the results are true when we replace $\mathbf{F}_\mathrm{ext}-\mathbf{F}_\mathrm{passive}$. 

\end{document}